\documentclass[12pt,preprint]{aastex}
\usepackage{times}
\usepackage{epsfig}

\shorttitle{Disk Evolution since $z\sim 1$}
\shortauthors{Brook et al.}

\begin{document}

\title{Disk evolution since $z\sim 1$ in a CDM Universe.}

\author{Chris B. Brook\altaffilmark{1}, Daisuke Kawata\altaffilmark{2,3}, 
        Hugo Martel\altaffilmark{1}, Brad K. Gibson\altaffilmark{3,4,5} , 
        Jeremy Bailin\altaffilmark{3}}

\altaffiltext{1}{D\'epartement de physique, de g\'enie physique et d'optique,
Universit\'e Laval, Qu\'ebec, QC, Canada  G1K 7P4}

\altaffiltext{2}{The Observatories of the Carnegie Institute of 
Washington, 813 Santa Barbara Street, Pasadena, CA 91101}

\altaffiltext{3}{Centre for Astrophysics, Univerrsity of Central Lancashire, Preston,
Pr1 2HE, United Kingdom}

\altaffiltext{4}{Laboratoire d'Astrophysique, Ecole Polytechnique Federale de Lausanne
(EPFL), Observatoire, CH-1290, Sauverny, Switzerland}

\altaffiltext{5}{Centre for Astrophysics \& Supercomputing, 
        Swinburne University,
	Hawthorn, Victoria, 3122, Australia}

\begin{abstract}
Increasingly large populations of disk galaxies are now being observed at 
increasingly high redshifts, providing new constraints on our knowledge of how 
such galaxies evolve. Are these observations consistent with a cosmology in 
which structures form hierarchically? To probe this question, we employ  
SPH/N-body galaxy scale simulations of late-type galaxies. We examine the
 evolution of these simulated disk galaxies from redshift 1 to 0, looking at 
the mass-size and luminosity-size relations, and the thickness parameter, 
defined as the ratio of scale height to scale length.  The structural 
parameters of our simulated disks settle down quickly, and after redshift $z=1$
the galaxies evolve to become only slightly flatter. Our present day simulated
galaxies are larger,  more massive, less bright, and redder than at $z=1$.
The inside-out nature of the growth of our simulated galaxies  reduces, and 
perhaps eliminates, expectations of evolution in the size-mass relation.

\end{abstract}


\section{INTRODUCTION}
Galaxies are major structural components of our Universe, yet their formation 
and evolution remain  an outstanding mystery of contemporary astrophysics. The 
latest generation of telescopes, including {\it Keck}, {\it Subaru}, the 
{\it Very Large Telescope}, {\it Gemini}, and the {\it Hubble Space Telescope},
provide snapshots of galaxies at various redshifts. The properties of disk 
galaxies out to $z=1$ have been the subject of several recent 
studies.  These properties include the Tully-Fisher relation 
(e.g. \citealt{vogt96,bm02,b04}), the luminosity-size, magnitude-size, and 
mass-size relations (e.g. \citealt{lilly98,simard99,bs02,b05}), the size 
distribution (e.g. \citealt{shen03,r04,F04}) and the disk ``thickness"
(\citealt{rdc03}, hereafter RDC, \citealt{elmegreen}).
 
The evolution of the luminosity-size and mass-size relations of disk galaxies 
remains a controversial issue. In particular, the interpretation of such 
evolution depends acutely on the selection biases of the surveys. Several 
studies have found an increase in the $B$-band surface brightness of 
$\sim 1$~mag~arcsec$^{-2}$, out to redshift 1 
(e.g. \citealt{schade,roche98,lilly98}). Taking into account selection effects
\cite{simard99} found minimal evolution with redshift, work that was supported 
by \cite{r04}, who found that the luminosity-size relation evolves by less than
0.4~mag~arcsec$^{-1}$ between $z=1.25$ and $z=0$. Yet these groups invoke a 
population   of high redshift, high surface brightness galaxies in their 
interpretation.  Different interpretations of completeness have  resulted in 
\cite{trujillo04} and \cite{b05}  finding an evolution in the rest frame 
$V$-band surface brightness of $\sim 0.8$ and $\sim 1.0$~mag~arcsec$^{-2}$ out 
to $z\sim 0.7$ and $z\sim 1.0$, respectively. \cite{b05} also find that disks 
at a given absolute magnitude are bluer, with lower stellar mass to light 
ratios at $z\sim1$, resulting in weak evolution in the relation between 
stellar mass and disk size. 

RDC selected 34 edge-on disk galaxies in the H{\sl ubble Deep Field} with
apparent diameters larger than 1.3\arcsec\ and unperturbed morphology.
They found an evolution between distant and local disk galaxies in their 
relative thickness or flatness, as characterized by the ratio of scale height
$h_z$ to scale length $h_l$. 
Their results indicate that disks at redshift $z\sim 1$ are 
smaller, in absolute value, than present day disk galaxies, and have a 
thickness ratio, $h_{\rm z}/h_{\rm l}$ larger by a factor $\sim 1.5$. 
support for such a thickening of disk galaxies at high redshifts also comes from \cite{elmegreen}, who inferred from their study of galaxies in the Hubble Ultra-Deep-Field that high redshift disk galaxies are thicker than local spirals by a factor of $\sim 2$.

Studies such as these are clearly increasing our knowledge of the evolutionary 
stages of populations of disk galaxies,  yet the path taken for galaxies to 
evolve between these stages remains unclear, and reliant on modeling. The 
general scenario of galaxy formation within the cosmological context, as 
envisioned by \cite{wr78}, involves extended dark halos of galaxies forming 
hierarchically by  gravitational clustering of non-dissipative dark matter.  
The luminous components form by a combination of the gravitational clustering 
and dissipative collapse.
A general framework for the formation of disk galaxies was outlined in 
\cite{fe80}.  In their model, a uniformly rotating, homogeneous protogalactic 
cloud of gas begins to collapse as it decouples from the Hubble flow.
This protogalaxy is endowed with angular momentum from tidal torques driven by 
surrounding structure (\citealt{p69}). It is assumed that the baryons destined 
to form the disk receive the same tidal torques as the dark matter before much 
dissipation occurs. If the collapse is smooth, then specific angular momentum 
is conserved and the gas forms a thin, rapidly rotating disk 
(e.g. \citealt{dss97,mmw98}). 

Detailed direct modeling  has highlighted that the above disk formation 
scenario is far from simple within a Cold Dark Matter (CDM) Universe . Complex 
baryonic processes involving multiple gas phases and energy feedback, for example from quasars, 
supernovae, and stellar winds, have proven difficult 
to model, yet these processes appear to be crucial in the formation and 
evolution of disk galaxies within the context of hierarchical structure 
assembly. Early simulations employed thermal feedback, following \cite{kg91}, augmented somewhat in \cite{nw93}, who added a variable fraction of supernova energy as kinetic energy. This type of feedback has been popular up until recent times (e.g. \citealt{sn02,abadi03}), even though it was known to be highly inefficient, with thermal energy radiating away over timescales shorter than the dynamical timescales \cite{katz92}. The effects of this inefficient feedback, in light of the the overcooling problem (\citealt{wr78}),  was the rapid formation of stars in the early collapsing halos. Subsequent mergers of these halos resulted in loss of angular momentum to the dark halo, resulting in disk galaxies being dominated by stellar halo components, and significantly deficient in angular momentum. \cite{weil} showed that the suppression of early radiative cooling in numerical simulations can result in more realistically sized disks. The problem was tackled in \cite{tc00} by turning off cooling for a fixed time in gas within the SPH smoothing kernel of a supernova event. Although lacking detailed insights into the multiphase gas processes involved in galaxy formation, these models succeeded in regulating star formation in small halos, and resulted in increased angular momentum in the simulated disk galaxy. In \cite{brooketal04a}, we simulated galaxies {\it{using identical initial conditions}} with these two different feedback methods. The thermal feedback following \cite{kg91}   resulted in galaxies which are dominated by a high mass, high metallicity spheroidal stellar  component, with a low mass disk component. By regulating star formation in low mass building blocks, the adiabatic feedback model, which follows \cite{tc00}, formed galaxies with a dominant disk component, and a low mass, low metallicity stellar halo.  Sommer-Larsen and collaborators have examined other implementations of feedback, and also shown the importance of regulating cooling at early epochs (\citealt{sl99,sl01,sl03}). Meanwhile, \cite{governato04} highlights the importance of resolution in simulating disk galaxies. Implementing these findings has allowed models such as \cite{robertsonetal04}, \cite{okamoto05} and \cite{governato05} to make progress in reproducing disk galaxies with properties approaching those of observed disks. 
  
Our chemo-dynamical models for simulating disk galaxies (which we will refer 
to as sGALS, to avoid confusion with real galaxies) have been successful in forming disk galaxies with a dominant young, metal-rich disk component, with a metal poor, old stellar halo, and an 
old, intermediate metallicity thick disk component. The sGALS have many 
features observed in the  Milky Way and local disk galaxies, including 
metallicities, abundances, ages,  and  color gradients of the stellar 
populations of the thin and thick disks, metallicities and abundances of the 
stellar halo, and the relation between luminosity and metallicity of the 
stellar halo (\citealt{brooketal04a}; \citealt{brooketal04b} [hereafter BKGF];
 \citealp{brooketal05} [hereafter BGMK]; \citealt{rendaetal05}).  
Here we report on the evolution of the disk component of such sGALS, 
specifically after redshift $\sim 1$. We deliberately follow the evolution 
of our sGALS at times after which they have settled into a disk morphology, 
and trace their luminosity, scale length, scale height, and central surface 
brightness. We hope to provide insight into how the disk component of 
late-type galaxies evolve in their relatively ``quiescent phase,'' during 
which mergers are only minor and star formation occurs almost exclusively in 
the disk regions. 

In this paper, we concentrate on the evolution of the size-mass and 
size-central surface brightness relations and the ``thickness'' of four 
simulated disk galaxies (sGALS), and what these tell us about interpreting 
observations.  We also examine the evolution of the magnitude and color of 
our simulated galaxies.
Our study will not allow direct comparison with the observational studies 
which compare distant galaxy populations to ``like'' local galaxy populations; 
it is not clear which galaxies are likely progenitors of today's galaxies. 
Yet, by providing evolutionary paths that disk galaxies can be expected to 
follow,  our study will provide insights into the interpretation of such 
observations, in particular in how to interpolate between the two populations. 

\section{THE CODE AND MODELS}

The  four sGALS studied here are the same as were studied in BGMK. They are formed 
using our galactic chemo-dynamical evolution code, {\tt GCD+}, which 
self-consistently models the effects of gravity, gas dynamics, radiative 
cooling, and star formation. Full details of {\tt GCD+} can be found in 
Kawata \& Gibson (2003). All galaxies  are formed with the 
{\it Adiabatic Feedback Model} from \cite{brooketal04a}.  In this model, gas within the SPH smoothing kernel of SNe II explosions is prevented from cooling. This adiabatic phase is assumed to last for the lifetime of the lowest mass star which ends as a SNe II, i.e. the lifetime of an 8-M star (100 Myr). In the AFM, the energy released by SNe Ia, which do not trace starburst regions, is fed back as thermal energy.

\begin{deluxetable}{cccc}
\tabletypesize{\footnotesize}
\tablecaption{Model Parameters for the 4 sGALS.\label{initial}}
\tablewidth{0pt}
\tablehead{\colhead{Galaxy} & \colhead{$z_c$} & 
\colhead{$M_{\rm tot} ({\rm M}_\odot)$} & \colhead{$\lambda$}}
\startdata
sGAL1 &  1.8  & 5$\times$10$^{11}$ & 0.0675 \\
sGAL2 &  1.9  & 5$\times$10$^{11}$ & 0.0675 \\
sGAL3 &  2.0  & 1$\times$10$^{12}$ & 0.0600 \\
sGAL4 &  2.2  & 5$\times$10$^{11}$ & 0.0675 \\
\enddata
\end{deluxetable}

We employ a semi-cosmological version of {\tt GCD+}, where  the initial 
conditions consist of an isolated sphere of dark matter and gas \citep{kg91}. 
This top-hat overdensity has an amplitude, $\delta_i$, at initial redshift, 
$z_i$, which is approximately related to the collapse redshift, $z_c$, by 
$z_c=0.36\delta_i(1+z_i)-1$ \citep[e.g.][]{pad93}.  Small scale density 
fluctuations are superimposed on each sphere, parameterized by $\sigma_8$,
 seeding  local collapse and subsequent star formation. Longer wavelength 
fluctuations are incorporated by imparting a solid-body rotation 
corresponding to a spin parameter, $\lambda=J|E|^{1/2}/GM_{\rm tot}^{5/2}$,  
to the initial sphere, where $J$ is the total angular momentum, $E$ the 
total energy, and $M_{\rm tot}$ is the total mass of the sphere.  
This simplified model allows us to run a suite of simulations at high resolution, whilst retaining the most important features of full cosmological simulations. 
The choice 
of large values of $\lambda$, and   initial conditions in which no major 
merger occurs in late epochs ($z<1$), ensures that disk sGALS are formed.
 We use an Einstein-deSitter CDM model;   parameters relevant to all sGALS
 include $\Omega_0=1$, baryon fraction, 
$\Omega_b=0.1$, $H_0=50\rm km\,s^{-1}\,Mpc^{-1}$, star formation 
efficiency, $c_*=0.05$, and $\sigma_8=0.5$. 
This choice was made merely for convenience. As we showed in BGMK,
the consequences of
using this cosmology with the initial conditions of this study, rather 
than the now standard $\Lambda$CDM cosmology, are negligible.
Essentially, prior to the redshift of collapse $z_c\sim2$, a $\Lambda$CDM
universe is almost identical to a Einstein-de~Sitter universe, while
after that redshift, the system is so dense that the background cosmology
becomes irrelevant. For a detailed proof, we refer the reader to BGMK. 
Other parameters varied slightly between the four 
models, but all were chosen to result in  
Milky-Way-type sGALS. 
Table~\ref{initial}, reproduced from BGMK, shows the values 
of total mass, $\lambda$, and $z_c$ for the four simulations. Along with these 
variations in parameters, different random seeds are incorporated in the 
initial conditions of the sGALS, creating evolutionary diversity in our 
sample. We employed 38911 dark matter and 38911 gas/star particles for 
each model.

\section{RESULTS}

\begin{figure*}
\epsfxsize=50.mm \plotone{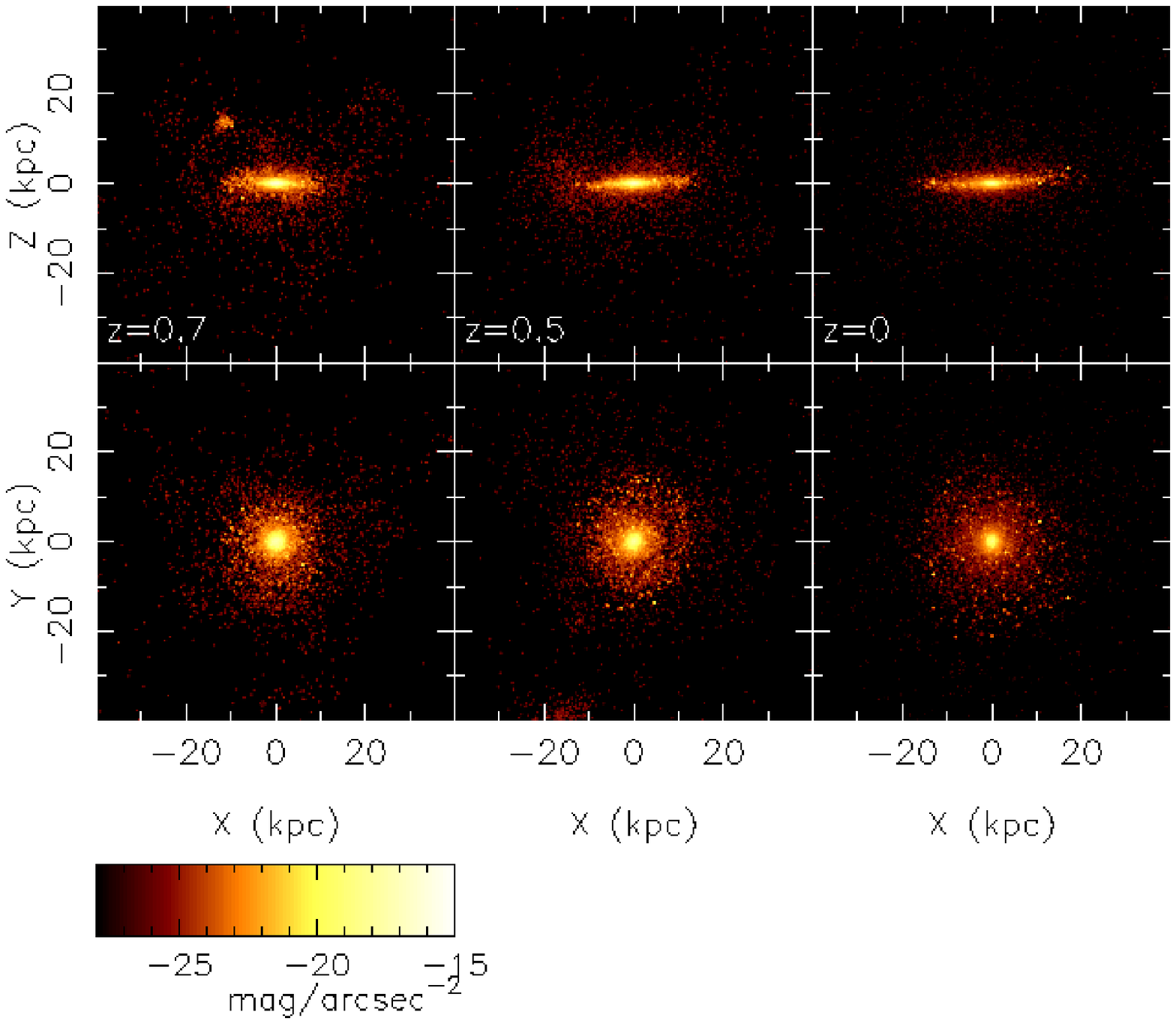}
\caption{Rest frame $B$-band luminosity plots  for sGAL1, plotted 
edge-on (upper panels) and face-on (lower panels) at redshifts $z=0.7$
(left panels), $z=0.5$ (middle panels), and $z=0$ (right panels).}
\label{SGAL1}
\end{figure*}

\begin{figure*}
\epsfxsize=50.mm \plotone{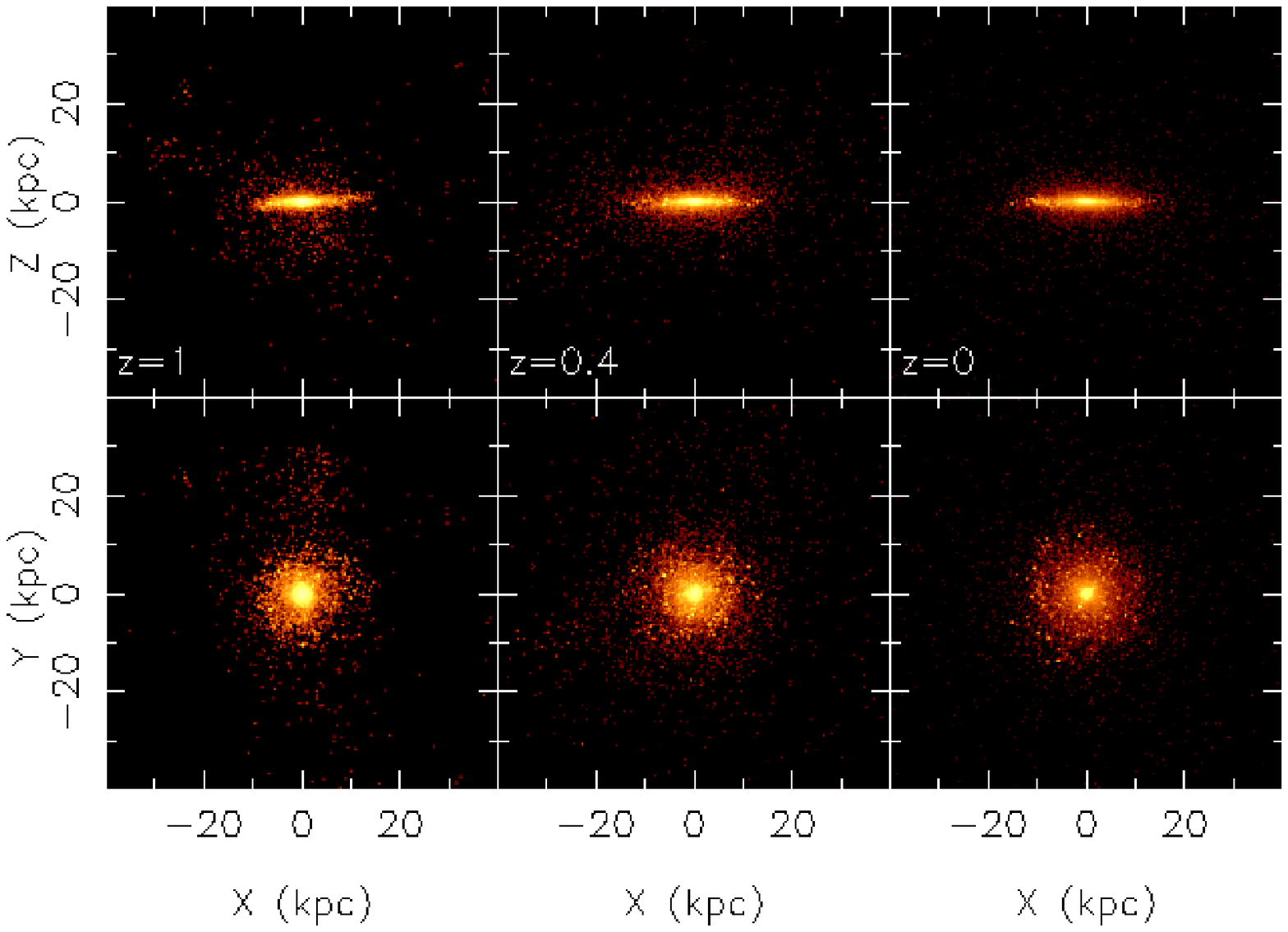}
\caption{Rest frame $B$-band luminosity plots  for sGAL2, plotted 
edge-on (upper panels) and face-on (lower panels) at redshifts $z=1$
(left panels), $z=0.4$ (middle panels), and $z=0$ (right panels).}
\label{SGAL2}
\end{figure*}

\begin{figure*}
\epsfxsize=50.mm \plotone{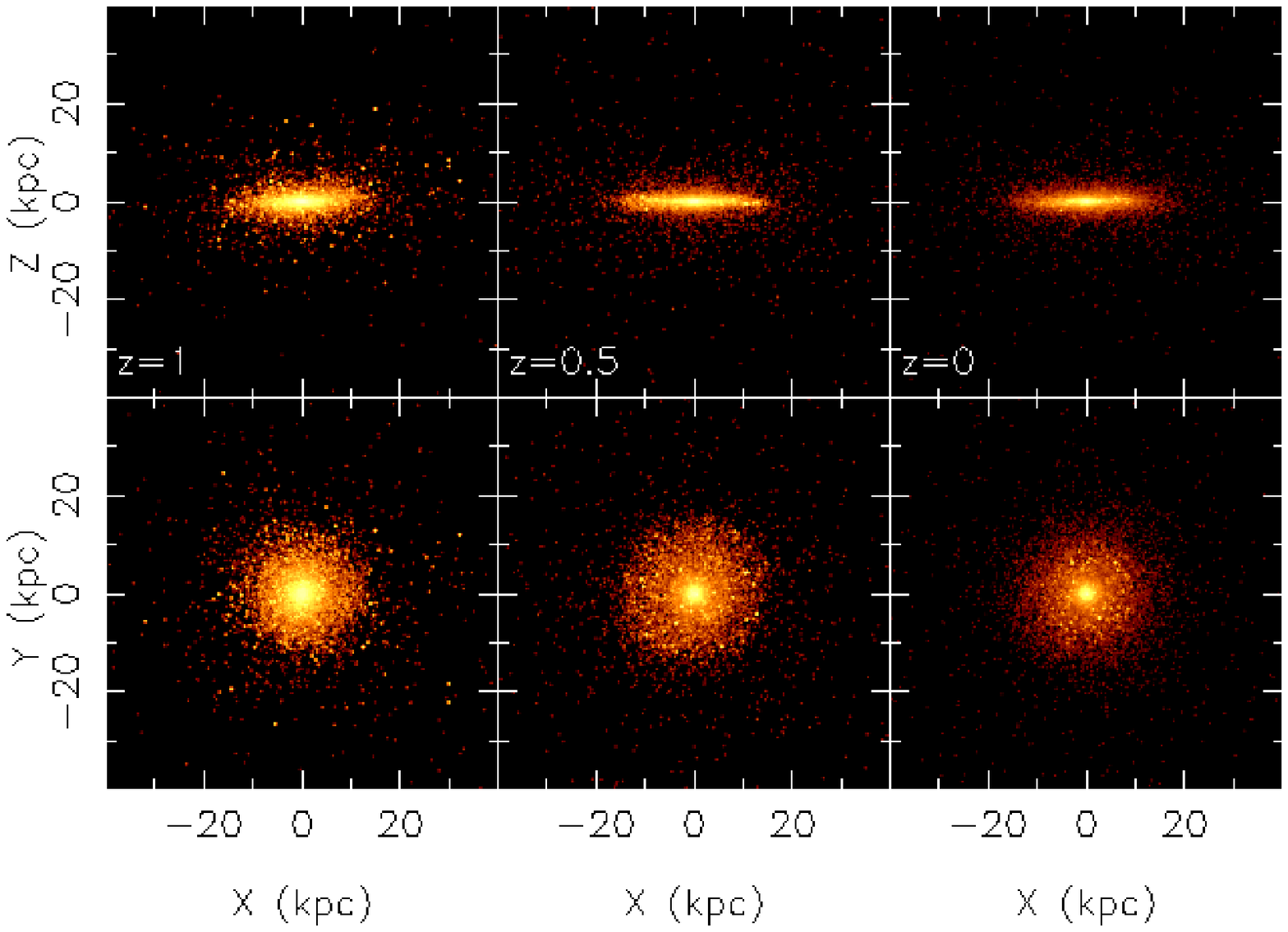}
\caption{Rest frame $B$-band luminosity plots  for sGAL3, plotted edge-on 
(upper panels) and face-on (lower panels) at redshifts $z=1$
(left panels), $z=0.5$ (middle panels), and $z=0$ (right panels).}
\label{SGAL3}
\end{figure*}

\begin{figure*}
\epsfxsize=50.mm \plotone{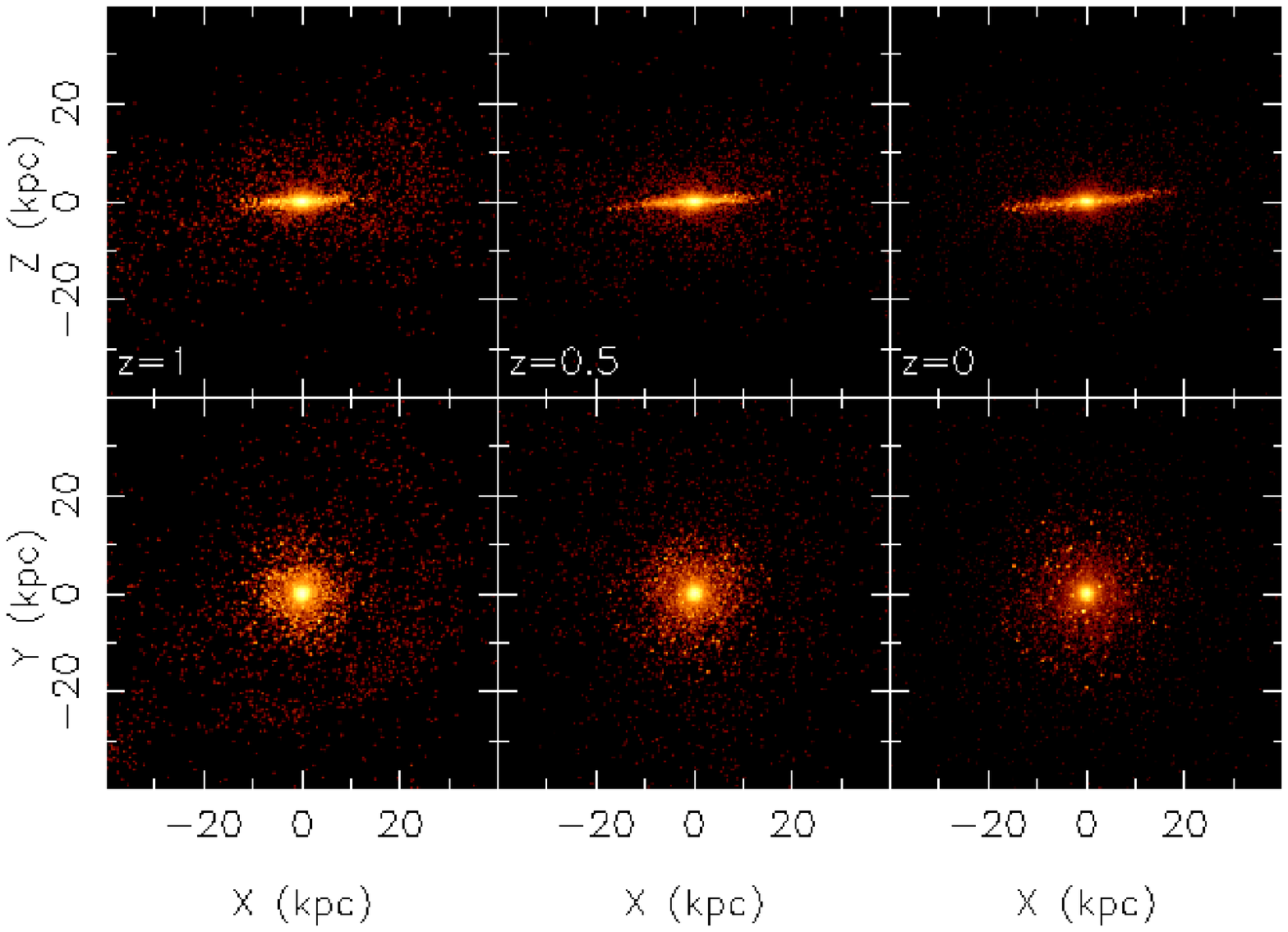}
\caption{Rest frame $B$-band luminosity plots  for sGAL4, plotted edge-on 
(upper panels) and face-on (lower panels) at redshifts $z=1$
(left panels), $z=0.5$ (middle panels), and $z=0$ (right panels).}
\label{SGAL4}
\end{figure*}

Our model results in the formation of late-type sGALS which have a dominant 
disk component, with a low mass, metal poor halo component (BKGF).
 Figure~\ref{SGAL1} shows the rest frame $B$-band surface brightness profile 
of sGAL1, both edge-on (upper panels) and face-on (lower panels), at three 
epochs: $z=0.7$ (corresponding to a lookback time of 7.3 Gyrs)  $z=0.5$ 
(5.7 Gyrs), and $z=0$. Figures~\ref{SGAL2}-\ref{SGAL4} show corresponding 
plots for sGALS 2-4, shown at  $z=1$ (corresponding to lookback time of 
8.5 Gyrs)  $z=0.5$ ($z=0.4$ for sGAL2, lookback time 4.8 Gyrs), and $z=0$. 
 These epochs are chosen to be: (a) after the thick disk formation epoch, 
when the old thin disk has begun to form; (b) at an intermediate time; and 
(c) the present, respectively. The different 
time chosen for the early epoch of sGAL1 ($z=0.7$) was taken as that 
simulated galaxy is undergoing a significant merger at $z=1$, and we need 
our sGALS to have a settled disk structure for a fair comparison with 
distant  galaxies which have been selected for their disk features. 
Similarly, a satellite is being accreted to sGAL2 at the intermediate 
epoch, $z\sim$0.5, so we chose a later intermediate time, $z=0.4$ for this 
sGAL.  The surface brightness is calculated by assuming that each star 
particle is a single stellar population, and using the SSPs of \cite{ka97}, 
with a simple Salpeter  initial mass function (IMF) parameterized by a power-law slope of 1.35 
and lower and upper mass of 0.1 and $60{\rm M}_\odot$, respectively. 
The photometric properties of 
sGALS discussed here are all intrinsic values, where we avoid discussing 
the effect of the dust absorption.

Total $B$-band magnitudes of the sGALS are found by integrating the 
luminosity densities displayed in these images. The results are shown 
in column~3 of Table~\ref{results}. We see that the galaxies have an 
average $M_B\sim -21.1$ at the earliest epoch (which we will refer to 
as $z\sim 1$), an average of $-20.6$ at the intermediate epoch ($z\sim0.5$) 
and $-19.7$ at $z=0$.
The higher luminosity at earlier times is largely due to the enhanced star 
formation at the epoch during which the thick disk forms (see BKGF), the 
epoch which precedes the ``early epoch'' ($z\sim1$)  analyzed in this 
study. The rate of star formation is lower during the more quiescent 
later epochs, resulting in evolution toward fainter magnitudes at the present. 

We repeat the above process in both the rest frame $V$ and $I$ bands, and 
derive $B-I$ and $V-I$ colors. The $B-I$ and $V-I$ colors  
are shown in columns~5 and~6 of Table~\ref{results}, respectively. The 
galaxies become redder during the relatively quiescent late epochs studied, 
from an average $B-I$ ($V-I$) of 1.1 (0.7) at $z$$\sim$1, to 1.5 
(0.9) at $z\sim0.5$, and 1.7 (1.0) at $z=0$. These redder colors are 
due to the effects of a decreasing star formation rate, aging stellar 
populations, and more metal rich stellar populations being born in the 
disk region.
The redder colors are not sufficient to overcome the decrease in $B$-band
luminosity; the sGALS also become fainter in the redder bands.
 
\begin{figure*}
\epsfxsize=50.mm \plotone{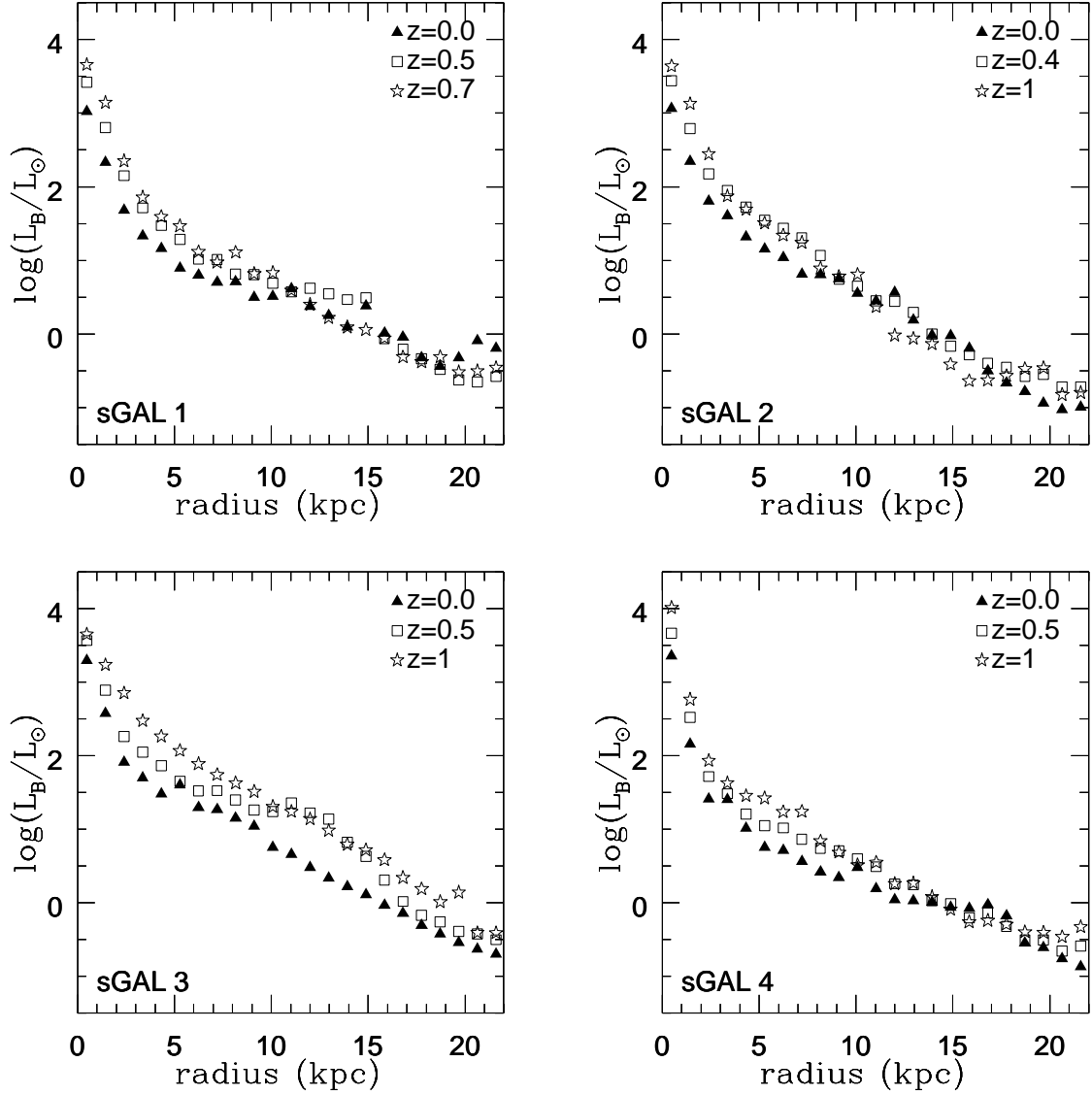}
\caption{Rest frame $B$-band luminosity (log scale) versus radius  for our 
4 sGALS,  
plotted at $z=1$ ($z=0.7$ for SGAL1, open stars), $z=0.5$ ($z=0.4$ for sGAL2, 
open squares) and  redshifts $z=0$ (solid triangles). 
Outside the bulge regions, the profiles are fitted reasonably by exponentials.}
\label{scalelengths}
\end{figure*}

The $B$-band surface brightness profiles for the four sGALS are plotted 
against radius in Figure~\ref{scalelengths}, at the three epochs shown in 
Figures~\ref{SGAL1}-\ref{SGAL4}. We use open stars for the early phase of 
evolution ($z\sim 1$), open squares for the intermediate phase ($z\sim0.5$), 
and solid triangles for $z=0$  in this and all ensuing plots in order to 
allow evolution of the properties of the disk sGALS to be easily followed. 
Outside the central bulge regions, the disk components can be approximated
reasonably by exponential fits. This allows us to derive disk scale lengths 
for each sGAL at each epoch. This is done in two stages. An approximate 
scale length $h_l^*$ is derived by fitting between 4 and 20 kpc. This allows 
us to then make fits between $1.5h_l^*$ (to exclude the bulge) and $4h_l^*$.  
Numerous tests verified that this technique gave robust results, which are 
shown column~9 in Table~\ref{results}. Scale lengths increase from an average 
of  2.9 at $z\sim1$ to an average of 3.5 at  $z\sim0.5$, and to 4.1 by 
$z=0$.  The implication is that the galaxies grow from  the ``inside-out.''

\begin{figure*}
\epsfxsize=50.mm \plotone{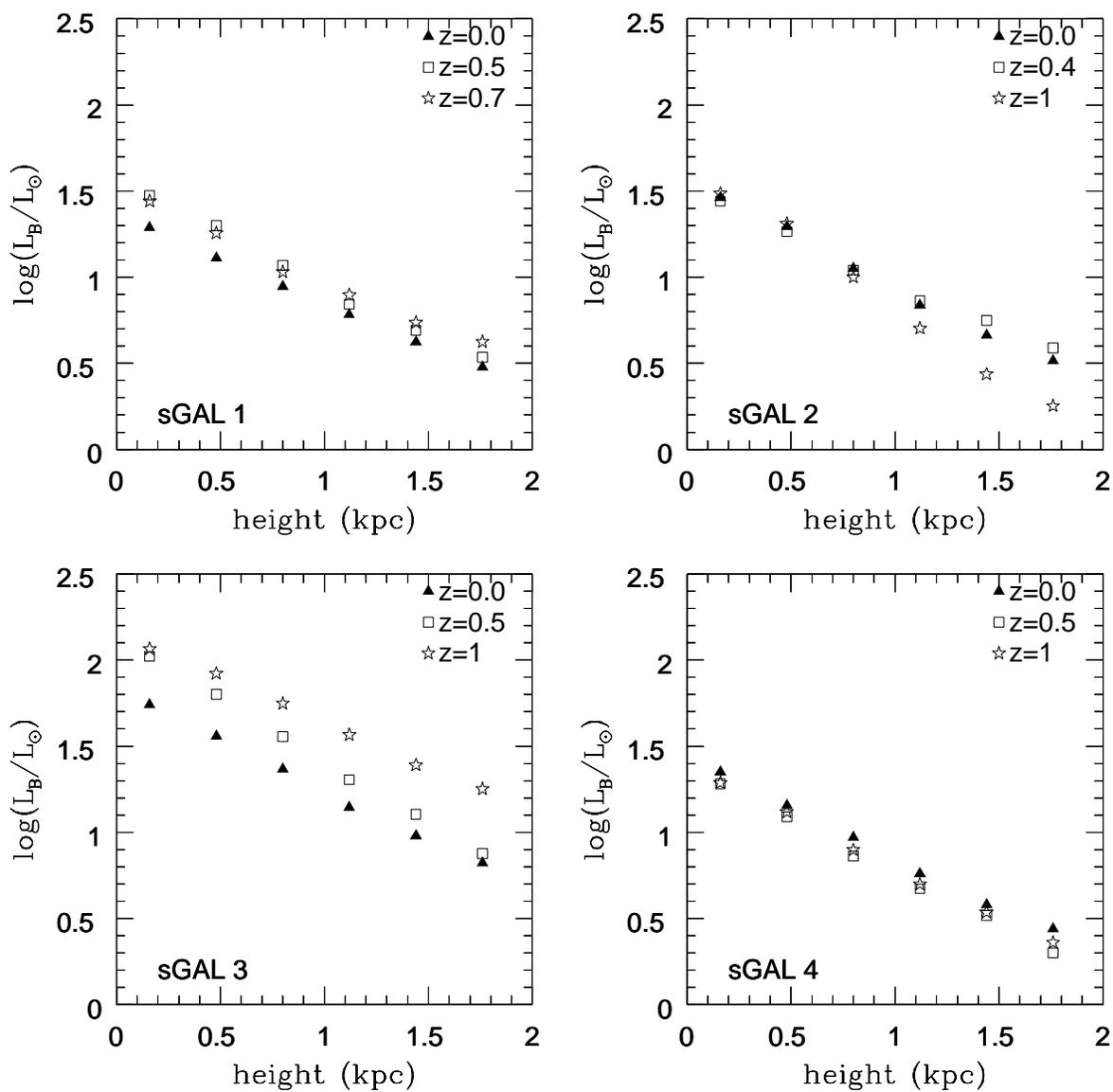}
\caption{Rest frame $B$-band luminosity (log scale) versus height above the 
disk plane for our four 
sGALS,  plotted at the same epochs, using the same symbols, as 
Figure~\ref{scalelengths}. The profiles are well fitted by exponentials.}
\label{scaleheights}
\end{figure*}

We plot the surface brightness profiles of the four sGALS perpendicular to the 
disk in Figure~\ref{scaleheights}. We exclude the bulge region when 
calculating how the surface brightness varies with distance from  the disk 
plane. Again, the profiles are well fitted with exponentials, and we fit 
between 0 and 2.5 scale heights (essentially out to 1.5 kpc in 
Figure~\ref{scaleheights}) using similar techniques as employed to derive 
scale lengths. The derived scale heights are shown in column~8 of 
Table~\ref{results}. Scale heights remain close to constant, with an 
average of $0.63$ at $z\sim 1$ to an average of $0.62$ at $z\sim 0.5$, and 
an average of $0.65$ at $z=0$. We derive the thickness of the disks, defined 
as the ratio of scale height to scale length, $h_z/h_l$, and show the 
results column~10 in Table~\ref{results}. On average, the galaxies are 
``thicker'' at the earlier epochs, by a factor of $1.4$ at $z\sim 1$.  
This is due to the no evolution of the scale hight and
 the increasing scale length with decreasing redshift.
 Our simulation does show little evolution of the scale height,
 because the thick disk forms first, followed by the thin disk
 formation, as discussed in BKGF. If the thick disk comes from heating of the thin disk, we should 
 expect an increasing scale height with decreasing redshift, and a slower evolution of the disk thickness. Further observations of the thickness evolution will help constrain thick disk formation scenarios. 
The evolution of the thickness in our sGALS (in the sense that the galaxies become thinner) is driven by the inside-out thin disk formation. At the 
intermediate epoch, the sGALS show a marginal increase in thickness, by a 
factor of $1.1$ at $z\sim 0.5$;  galaxies settle rapidly to disk morphologies 
after their last ``significant'' merger (i.e. a merger likely to alter the 
morphology of the galaxy).  
 
 Numerical heating is an issue when considering vertical structure in numerical simulations, but we have reason to believe that our results are not  affected significantly. Previous studies at similar resolution to ours have shown numerical vertical heating to be relatively low (\citealt{font,mayer}).  Figure 2 of BGMK shows the vertical velocity dispersion versus age for four sGALS in this study, which demonstrates relatively little vertical heating during the disks' quiescent period; in the order of 20\% over the last $\sim 8$ Gyrs. Even if this were totally attributed to numerical heating (unlikely as processes such as spiral structure and substructure heating will contribute), our results will not be greatly affected, because the vertical   scale height is proportional to the square root of the velocity  dispersion, i.e $h_z\propto \sqrt{\sigma_z}$ (e.g. \citealt{bottema}). Our measured scale height at $z=0$ will be, in the worst case, overestimated by $\sim$10\%. As far as relevance to our study, the  important point  that we note is that any artificial heating will not affect our results {\it{qualitatively}}. In the absence of such effects, our conclusions that our simulated disk galaxies become {\it{flatter}} ($h_z/h_l$ decreases) will hold. Our {{\it quantitative}} estimates would then be underestimating this effect. Having said this, higher resolution studies will increase the reliability of  the {\it{quantitative}} estimates
 
\begin{figure*}
\epsfxsize=50.mm \plotone{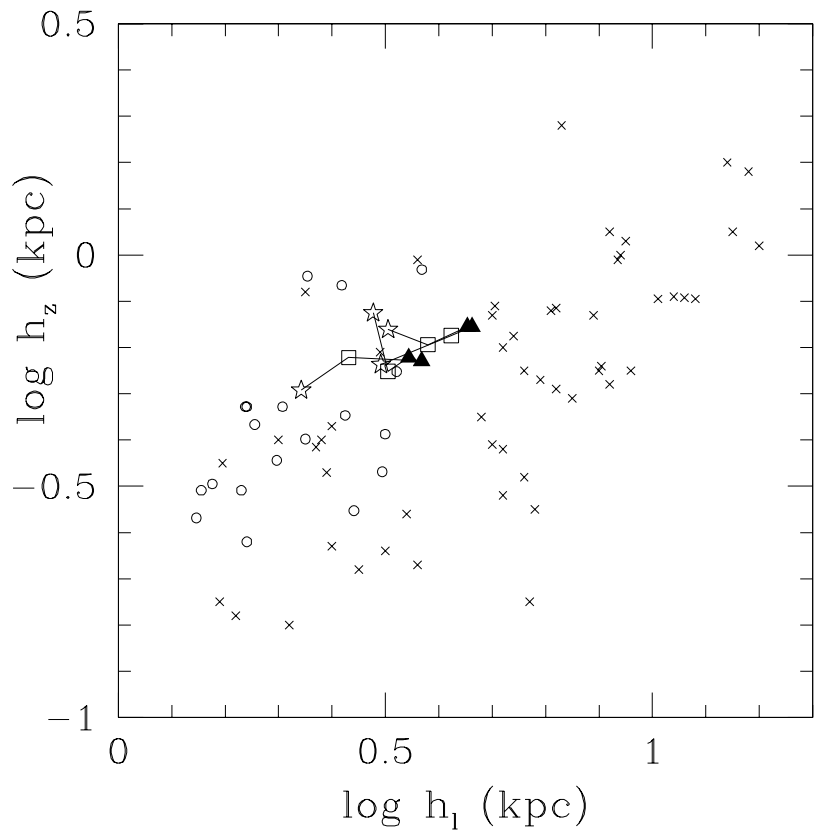}
\caption{Log (scale length) versus log (scale height) for our four sGALS, with 
symbols responding to the three epochs as defined in 
Figures~\ref{SGAL1}--\ref{SGAL4}.  Individual sGALS are joined by lines. Also plotted  are observations of local 
disk galaxies from \cite{sd00} (crosses),
and observations  of edge-on disk galaxies at $z\sim 1$ from RDC 
(open circles). Our sGALS evolve primarily by increasing scale length, with 
little evolution in scale height;  disk growth is inside-out.}
\label{hzhl}
\end{figure*}

 The way that the structural parameters, scale length and scale height, 
evolve relative to one another can be ascertained from  Figure~\ref{hzhl}. 
Also plotted  as crosses are observations of local disk galaxies from 
Schwarzkopf \& Dettmar (2000). Plotted as open circles are  observations  
of 34 edge-on disk galaxies at $z$$\sim$1, taken from RDC. The 
$z\sim1$ disk galaxies
of RDC are found to be smaller and thicker than local galaxies, 
as determined by previous studies, such as \citet{sd00}, \citet{dG98}, 
\citet{K02} and \citet{bm02}. Our sGALS evolve in this plot primarily by 
increasing in scale length, with less evolution in scale height, resulting 
in galaxies evolving across this diagram, from left to right.

 Central surface brightness, again in the rest frame in the $B$-band, is 
derived by extrapolating the exponential fits from Figure~\ref{scalelengths} 
to $r=0$. Results are shown in column~7 of  Table~\ref{results}. We find that 
our sGALS are brighter at earlier epochs, with an average $\mu_B=21.0$ at 
$z\sim 1$, compared with $\mu_B=21.8$ at $z\sim 0.5$, and $\mu_B=22.7$ at 
$z=0$.  This is related to the higher star formation rates at the earlier 
epochs in our sGALS, as shown in Figure~5 of BGMK.  This is consistent with 
observations, which  point to star formation rates being higher at earlier 
epochs (e.g. \citealt{madau,hammer,flores,juneau}). We plot the $B$-band 
central surface brightness  against the flatness parameter, $h_l/h_z$
(the inverse of the thickness parameter), in Figure~\ref{muBhlhz}.  
 The sGALS have higher surface brightness and
are thicker at earlier times, closer to the merging epochs when the thick 
disks formed. As the thin disk grows larger in the more quiescent later 
stages of their evolution, the sGALS become less bright, and also thinner.
At each epoch, it is also apparent that brighter sGALS are thicker, although 
this trend is not as strong as the trend with time, because  the sGALS we 
have studied are all relatively quiescent  throughout the epoch during which 
evolution is followed in this study.
This supports the conclusions of RDC, who plotted the relation 
between $h_l/h_z$ and central surface brightness of $z\sim 1$ galaxies 
(their Fig~.7), and found that the brightest objects have the thickest disk. 
They speculated that enhanced star formation and enhanced thickness are the 
results of ongoing interactions and mergings.
 
We derive the stellar mass $M_*$ of each sGAL at each epoch using the friends-of-friends algorithm, and confirm that sensitivity to our choice of linking 
length is insignificant. Results are displayed in column~4 of 
Table~\ref{results}.  The inside-out nature of the growth of the thin disk 
in our sGALS is apparent in the relationship between the stellar mass and 
scale length, shown in Figure~\ref{Masshl}. The mass and size of the sGALS 
grow together. Although we do not want to overemphasize the quantitative 
nature of our  results, using just four relatively homogeneous  L$_*$ sGALS, 
we note that the ratio $M_*/h_l$ does not evolve, with the average at each 
epoch staying remarkably close to $10^{10}{\rm M}_{\Sun}\,\mathrm{kpc}^{-1}$.  
Qualitatively, it is clear that the thin disks  of our sGALS increase in size as 
they grow in mass, and the disk evolves along the mass scale length relation.

\begin{deluxetable}{cccccccccc}
\tabletypesize{\footnotesize}
\tablecaption{Derived parameters of our four sGALS at present ($z=0$), intermediate ($z\sim 0.5$) and early ($z\sim 1$) epochs, in columns: (3) $B$-band magnitude (4) Stellar Mass (5) $B$-$I$ color (6)  $V$-$I$ color (7) $B$-band central surface brightness (8) scale height (9) scale length (10) thickness parameter }
\tablewidth{0pt}
\tablehead{\colhead{redshift} & \colhead{SGAL} & \colhead{$B$(mag)} & \colhead{$M_*$ (M$_\odot$)} & \colhead{$B$$-$$I$} & 
\colhead{$V$$-$$I$}& \colhead{$\mu_B$(mag/arcsec$^{2}$)} & \colhead{$h_z^B$(kpc)} & \colhead{$h_l^B$(kpc)}& \colhead{$h_z^B/h_l^B$}
}
\startdata
$z=0$     & sGAL1 & -19.5 & 2.9$\times$10$^{10}$ &1.7 & 1.0 & 23.3 & 0.70 & 4.5 & 0.16\\
$z=0$     & sGAL2 & -19.6 & 3.0$\times$10$^{10}$ &1.7 & 1.0 & 22.1 & 0.59 & 3.7 & 0.16\\
$z=0$     & sGAL3 & -20.2 &  6.6$\times$10$^{10}$&1.8 & 1.1 & 21.4 & 0.60 & 3.5 & 0.17\\
$z=0$     & sGAL4 & -19.6 &  3.6$\times$10$^{10}$&1.7 & 1.0 & 23.9 & 0.70 & 4.6 & 0.15\\
\tableline
$z=0$     & mean      & -19.7 & 4.0 $\times$10$^{10}$&1.7 & 1.0 & 22.7 & 0.65 & 4.1 & 0.16\\
\tableline
$z=0.5$  & sGAL1 & -20.3 & 2.4$\times$10$^{10}$&1.5 & 0.9 & 22.2 & 0.64 & 3.8 & 0.17\\
$z=0.4$   & sGAL2 & -20.6 &2.4 $\times$10$^{10}$&1.3 & 0.8 & 21.0 & 0.60 & 2.7 & 0.22\\
$z=0.5$  & sGAL3 & -21.0 & 5.9$\times$10$^{10}$&1.5 & 0.9 & 20.6 & 0.56 & 3.2 & 0.18\\
$z=0.5$  & sGAL4 & -20.3 & 3.2$\times$10$^{10}$&1.5 & 0.9 & 23.2 & 0.67 & 4.2 & 0.16\\
\tableline
$z\sim 0.5$  & mean      & -20.6 & 3.5$\times$10$^{10}$&1.5 & 0.9 & 21.8 & 0.62 & 3.5 & 0.18\\
\tableline
$z=0.7$   & sGAL1 & -20.9 &2.1$\times$10$^{10}$& 1.1 & 0.7 & 21.8 & 0.69 & 3.2 & 0.22\\
$z=1$     & sGAL2 & -20.9 & 1.4$\times$10$^{10}$&0.9 & 0.6 & 20.7 & 0.51 & 2.2 & 0.23\\
$z=1$     & sGAL3 & -21.6 & 5.0$\times$10$^{10}$&1.3 & 0.8 & 20.2 & 0.75 & 3.0 & 0.25 \\
$z=1$     & sGAL4 & -21.1 & 2.7$\times$10$^{10}$&1.2 & 0.8 & 21.6 & 0.58 & 3.1 & 0.22\\
\tableline
$z\sim0.9$ & mean   & -21.1 &2.8$\times$10$^{10}$ &1.1 & 0.7 & 21.0 & 0.63 & 2.9 & 0.23\\ 
\tableline
\enddata
\label{results}
\end{deluxetable}

\section{SUMMARY AND CONCLUSION}

The evolution of individual galaxies cannot be observed directly, but 
lookback studies are providing 
increasing information on properties of  populations of ``like'' galaxies, 
out to redshift 1 and beyond.  These studies remain difficult, 
in large part because it is unclear how to select which galaxies should be
compared at each epoch.
In addition, there are large uncertainties involved in
correcting for dust and cosmological dimming,
especially in  the case of edge-on disk galaxies.

Several recent studies have compared the observed evolution of disk galaxy 
properties with  semi-analytic and infall type models. For example, 
\cite{b05} and \cite{trujillo05} show that one of the most basic 
assumptions of the semi-analytic approach, where disk size scales in 
proportion to the virial radius of the dark halo \citep{mmw98}, is at 
odds with observations; the observed size evolution is weaker than 
the predicted $R\propto H(z)^{-2/3}$, where $H(z)$ is the Hubble parameter as a function of redshift. The infall model of \cite{bs02} 
assumes a star formation rate at each radius and time, according to the 
local density (\citealt{Schmidt}), and the infall of metal-free gas. 
This model is able to match observations of the size evolution, at given 
luminosity, out to $z\sim1.7$ (\citealt{trujillo05}). In order to explain the evolution of disk galaxies, the necessity of accounting for effects such as mass build up, star formation, energy 
feedback, and gas infall self-consistently within a CDM hierarchical framework,  motivates us to confront observations with direct numerical simulations. 

\begin{figure}
\epsfxsize=50.mm \plotone{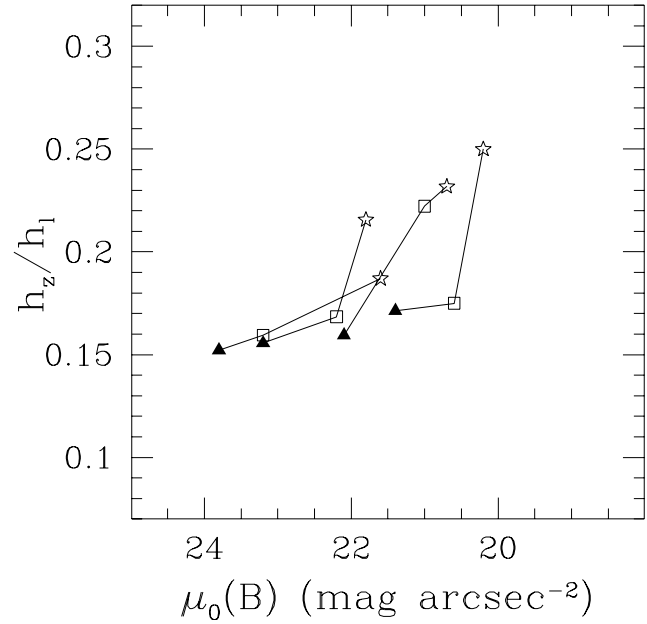}
\caption{The thickness of the sGALS, as defined by the ratio of scale height 
to scale length, plotted against the $B$-band rest frame central surface 
brightness. We show the early (open stars), intermediate (open squares) and 
present (solid triangles) epochs, as defined in 
Figures~\ref{SGAL1}--\ref{SGAL4}. Individual sGALS are joined by lines. The sGALS have 
higher surface brightness and are thicker at earlier epochs. At each epoch, 
it also is apparent that 
higher surface brightness sGALS are thicker, although this trend is not as 
strong as the trend with time.  }
\label{muBhlhz}
\end{figure}

\begin{figure*}
\epsfxsize=50.mm \plotone{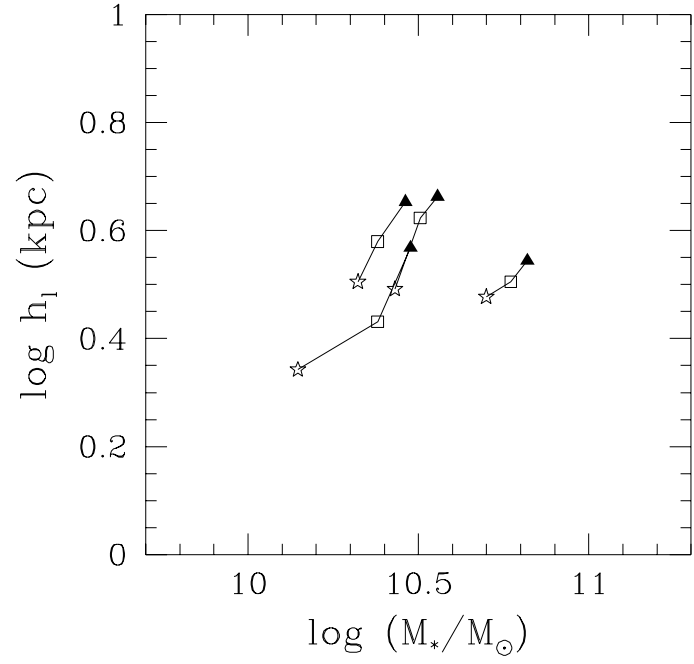}
\caption{Log of the stellar mass of the sGALS against log scale length, 
with symbols representing the epochs as shown in previous diagrams.  Individual sGALS are joined by lines. 
As the sGALS become more massive, they also become larger, growing from 
the inside-out.}
\label{Masshl}
\end{figure*}

We examine four simulated disk galaxies (sGALS), and do so in an idealized 
manner, free from the difficulties posed in observation.  Our sGALS form 
thick disks at high redshift ($z\sim 1.5-2$) during a period of mergers of 
gas rich ``building blocks'' (BKGF), and rapidly settle into a disk galaxy 
morphology by $z\sim1$. We examine their subsequent evolution to the 
present, comparing various properties of the galaxies at high ($\sim1$), 
intermediate ($\sim0.5$), and zero redshift, where we find clear trends. 
Our main results are:

\begin{itemize}
\item Scale heights of disk galaxies show no evolution after $z\sim 1$.
 Scale lengths increase slowly, resulting in flatter disks at later times.
 This is consistent with the observations of RDC, who find that 
disk galaxies at $z\sim 1$ are smaller than present-day galaxies, and have 
a thickness ratio, $h_z/h_l$ larger by a factor of $\sim 1.5$.

\item The  ``inside-out'' growth of disks is reflected in the
scale length increasing roughly proportional to the stellar mass,
resulting in little to no evolution of the mass-size relation.
This is consistent with interpretations of \cite{trujillo04,b05,trujillo05}.
Previously, few  direct numerical studies have probed these issues. In a 
study of sGALS  of various morphologies in a fully self-consistent 
$\Lambda$CDM simulation, \cite{sl03} examined two disk galaxies, and found 
that while one clearly evolved inside-out, the other evolves 
outside-in.  \cite{Samland} examine the build up of disk galaxies 
through a spherical gas infall model, perhaps  an idealization of the 
quiescent period of disk galaxy formation, and finds that disks form 
inside-out.

\item Disk sGALS become more massive, less bright, and redder between 
redshifts $z\sim1$ and 0.
 The evolution of the luminosity of disk galaxies remains observationally 
controversial, with selection effects largely determining interpretations 
(for a discussion see \citealt{b05}).  The mass build up in this epoch is 
relatively quiescent, with aging stellar populations and high metallicity 
stars born in the disk accounting for the galaxy reddening.

\item Rapid periods of star formation due to mergers and interactions can 
result in both thickening of the disk, and galaxy brightening 
(see Figure~\ref{muBhlhz}).
 This is consistent with the results of RDC, as well as local 
observations which show that 
interacting galaxies are thicker (\citealt{rc97}), and that  a correlation 
exists between central surface brightness and ``thickness,'' in the sense 
that higher surface brightness galaxies are thicker (\citealt{b00,bm02,bk04}).
\end{itemize}

The quantitative results of our study are summarized in Table~\ref{results}. 
These results are subject to uncertainties inherent in our models, such as 
star formation and feedback prescriptions. Also, our simulations do not 
cover the range of galaxy masses, sizes, morphologies, and histories 
of the large observed populations.
We therefore feel that more modeling of a greater population of high- and 
low-redshift sGALS within a fully cosmological simulation is required in 
order to make more meaningful quantitative predictions. Having said this, 
clear  trends have emerged in  this study of the  evolution of the 
properties of disk sGALS formed in a hierarchical, Cold Dark Matter Universe.  

\acknowledgments 
We thank Agostino Renda for his help. 
Simulations were performed on Swinburne University Supercomputer, those of the Australian Partnership for Advanced Computing, and on ``Purplehaze,'' the Supercomputer facility at Universit\'{e} Laval. CB and HM thank the Canada Research Chair program and NSERC for support. BKG, DK \& JB acknowledge the financial support of the Australian Research Council through its Discovery Project program. DK acknowledges the financial support of the Japan Society for the Promotion of Science, through a Postdoctoral Fellowship for research abroad.

\end{document}